\documentclass[runningheads]{llncs}
\usepackage[T1]{fontenc}
\usepackage{graphicx}
 \usepackage{multirow}
\usepackage{aecompl}
\usepackage{changes}
\usepackage{changepage}
\usepackage{amsmath}
\usepackage{subcaption}
\usepackage{float}
\begin{document}
\title{Travel Time Based Task Mapping for \\NoC-Based DNN Accelerator}
%
%

\author{Yizhi Chen\inst{1}\orcidID{0000-0001-8488-3506}\and Wenyao Zhu\inst{1}\orcidID{0000-0002-4911-0257}  \and Zhonghai Lu\inst{1}\orcidID{0000-0003-0061-3475}}

\authorrunning{Y. Chen et al.}

%
\institute{KTH Royal Institute of Technology, Stockholm, Sweden \email{\{yizhic,wenyao,zhonghai\}@kth.se}}
\maketitle              

\newcommand\Fig[1]{\textbf{Fig.}~\ref{#1}}
\newcommand\fig[1]{\textbf{Fig.}~\ref{#1}}
\newcommand\Tab[1]{\textbf{Tab.}~\ref{#1}}
\newcommand\tab[1]{\textbf{Tab.}~\ref{#1}}
\newcommand\Equ[1]{\textbf{Eq.}~(\ref{#1})}
\newcommand\equ[1]{\textbf{Eq.}~(\ref{#1})}
\newcommand\Sect[1]{\textbf{Sec.}~\ref{#1}}
\newcommand\sect[1]{\textbf{Sec.}~\ref{#1}}
\newcommand\Refs[1]{\textbf{Ref.}~\cite{#1}}
\newcommand\refs[1]{\textbf{Ref.}~\cite{#1}}

\begin{abstract}
\par 
Network-on-Chip (NoC) based architectures are recently proposed to accelerate deep neural networks in specialized hardware. Given that the hardware configuration is fixed post-manufacture,  proper task mapping attracts researchers' interest. We propose a travel time based task mapping method that allocates uneven counts of tasks across different Processing Elements (PEs).  This approach utilizes the travel time recorded in the sampling window and implicitly makes use of static NoC architecture information and dynamic NoC congestion status. Furthermore, we examine the effectiveness of our method under various configurations, including different mapping iterations, flit sizes, and NoC architectures.  Our method achieves up to 12.1\% improvement compared with even mapping and static distance mapping for one layer. For a complete NN example, our method achieves 10.37\% and 13.75\% overall improvements to row-major mapping and distance-based mapping, respectively.  While ideal travel time based mapping (post-run) achieves 10.37\% overall improvements to row-major mapping, we adopt a sampling window to efficiently map tasks during the running, achieving 8.17\% (sampling window 10) improvement.
\end{abstract}
%
\section{Introduction}
 \par Deep Neural Networks (DNNs) have emerged as one of the hottest topics in both academic and industry circles \cite{xue2022aome,kwon2020maestromapping}. Due to the high computational complexity of DNN, hardware acceleration abstracts significant interest among researchers. 
 
 \par Notably, Network-on-Chip-based (NoC-based) accelerators \cite{shao2019simba,zhu2024activation} offer a balance between performance and flexibility, establishing a new paradigm in design approaches. Furthermore, NoCs have demonstrated remarkable potential in facilitating efficient on-chip data communication for operating neural networks \cite{xue2022aome}.
\par With the specific hardware solutions being fixed after manufacturing, the subsequent challenge is efficiently mapping a DNN layer to the NoC platform \cite{kwon2020maestromapping}.  The prevalent method, even mapping in \cite{chen2019eyerissV2}, entails distributing tasks equally across all available Processing Elements (PEs) in a single iteration and repeating this process until task allocation is complete. However, even mapping does not consider the variance between different PEs, such as the distance to memory, and the run-time status of the NoC, leading to an unbalanced workload and time consumption.

Our research aims to enhance the efficiency of accelerators by mapping different numbers of tasks among PEs, due to their inherent differences and varying running status. By allocating fewer tasks to slower PEs, we aim to balance the NoC workload, minimize idle time, and thereby improve overall performance.

The contributions of our work are summarized as follows:
\begin{itemize}
 \item  We propose a travel time based mapping strategy to unevenly distribute tasks to PEs and reduce time consumption by balancing workload in NoC-based DNN accelerators. 
\item  We investigate the impact of different NN layers, communication protocols, and hardware platforms by analyzing the performance of a layer under various conditions of mapping iterations, packet size, and NoC architecture. Additionally, we evaluate our method on a complete DNN model, LeNet\cite{lenet}.
 \item  We propose an on-the-fly travel time based mapping with a sampling window. We discuss the influence of various sampling window lengths.
\end{itemize}
\par  In \sect{relatedwork}, we discuss the related work. In \sect{Task mapping bg}, we introduce the state-of-the-art task mapping methods and then present our travel time based mapping method in \sect{Task mapping method}. In \sect{sectexperiments}, we show the experimental results to validate our approach. In \sect{sectconclusion}, we conclude this paper.

\section{Related Work}
\label{relatedwork}
\par   Strategically placing application tasks on processing cores has emerged as a crucial component in the design of NoC-based MPSoCs (Multi-processor System-on-Chip) for optimal performance \cite{PerformanceEvaluationGeneralMapping}. Many works focus on non-DNN tasks including Best Neighbor mapping \cite{bestneighbormpeg4} for MPEG-4 and multiwindow display (MWD) tasks, \cite{stuijk2008resource} for synthetic traffic and MPEG-4 applications, L-shape isolated (Liso) mapping \cite{sadeghi2019toward} for PARSEC and SPLASH-2 benchmarks. The mapping of such applications and benchmarks usually needs to consider diverse task dependencies, which leads to non-regular traffic patterns in NoC.

\par   DNNs mapping deals with significantly different data traffic patterns from that in conventional MPSoCs \cite{neunoc}. GAMMA \cite{GAMMA} introduces a specialized genetic algorithm performing a search in the massive mapping space.  Autonomous Optimal Mapping Exploration (AOME) \cite{xue2022aome} leverages two reinforcement learning algorithms to efficiently explore optimal hardware mapping. ZigZag \cite{zigzaguneven} extends the normal design space exploitation by introducing uneven mapping opportunities. These methodologies, prioritizing the exploration, are time intensive.
\par Dense mapping \cite{densemapping} incorporates an input-sharing mechanism to facilitate the reuse of input data, thereby conserving resources. Similarly, Neu-NoC \cite{neunoc} aims to minimize redundant data traffic within neuromorphic acceleration systems and improve data transfer capabilities between adjacent layers. Configuring their nodes to transmit data to other computation nodes executing adjacent layers instead of directly to memory nodes makes the hardware complex. \cite{russo2023memorylinearmapping} uses integer linear programming to map the DNN layer, but it still maps the tasks evenly. 
\par Load balancing, aiming to ensure equitable utilization in NoC, could be used for DNN tasks. \cite{mondal2022gnnie} develop a static weight redistribution method for graph neural network inference using a PE array. Their approach focuses on maximizing the reuse of cached data. 
Work-stealing \cite{blumofe1999scheduling} is a dynamic load-balancing strategy that allows idle PEs in the network to actively seek tasks from busier PEs to optimize resource utilization and improve overall performance. Girao \textit{et al.} \cite{girao2013exploring} explores work-stealing mapping policies to determine whether a task migration between the cores should take place.  It is flexible but frequently collecting the real-time workload status from other PEs is a large overhead for DNN tasks. time based partitioning is also introduced in \cite{shao2019simba}, but it only considers communication latency and requires an initial execution. Additionally, it does not explore the effects of varying configurations on the partitioning.

\par Our method leverages travel time within a sampling window to unevenly distribute DNN tasks. It effectively balances the workload using a simple control algorithm and does not require an extra run.  

\section{Task Mapping Methods}
\label{Task mapping bg}

 \subsection{DNN Tasks and Mapping to NoC}
   \begin{figure}[htb]
      \centering
    \includegraphics[width = 10 cm]{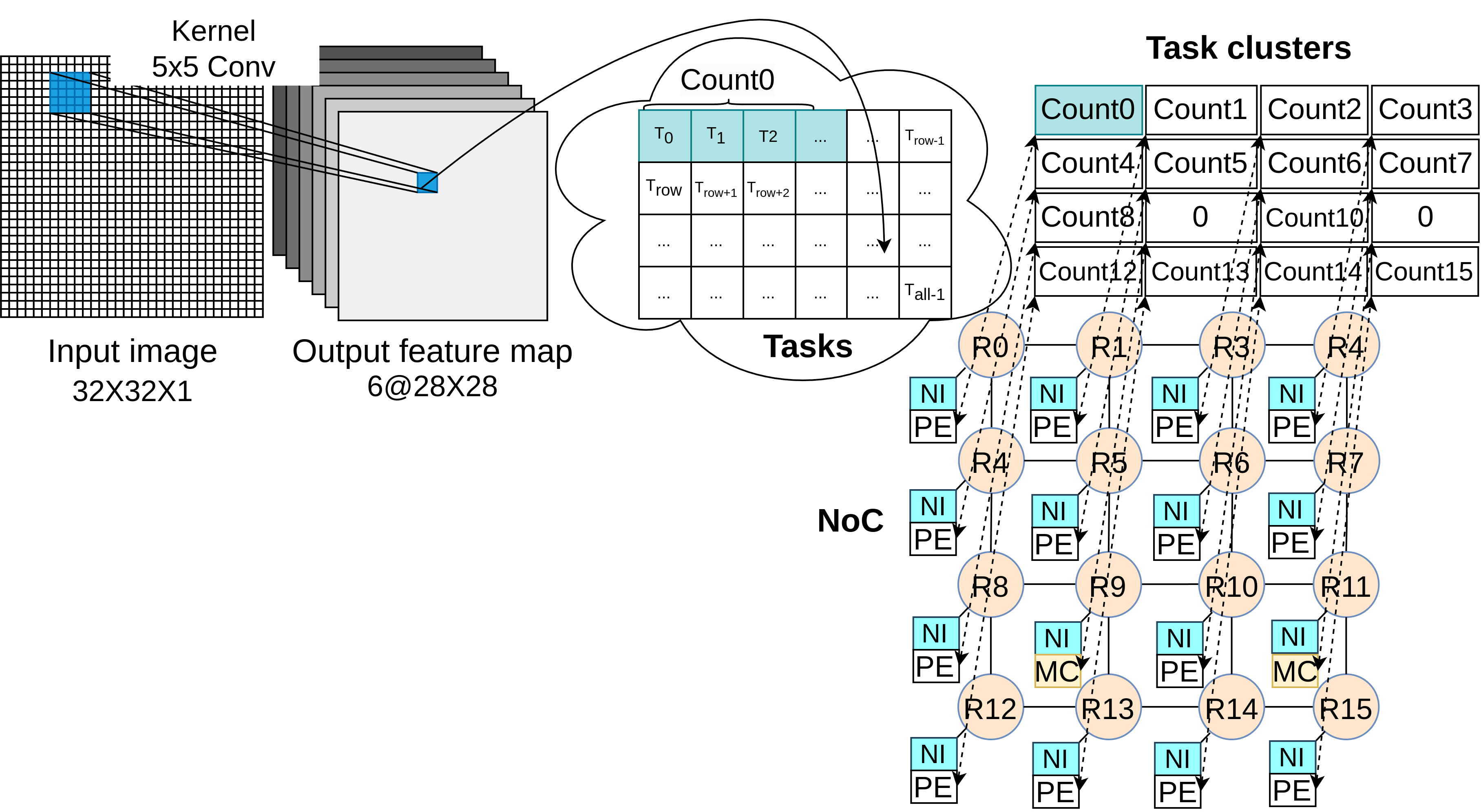}
      \caption{Tasks mapping}
      \label{fig:taskmapping_horizonal}
  \end{figure}
We introduce the tasks to be mapped from a DNN model on the left side of \fig{fig:taskmapping_horizonal}, using the first layer of LeNet as an example. This layer processes a 32x32 padded input image through a 5x5 kernel convolution. This convolution operation constitutes a computation task and yields a pixel in the output feature map. 
\par Mapping to NoC involves assigning these computation tasks in the cloud-shaped area in \fig{fig:taskmapping_horizonal} to specific hardware resources.  While a 4x4 NoC serves as an example, mapping is to allocate tasks to 16 clusters. For example, the first cluster on the right side, highlighted in blue, contains $Count0$ tasks, correspondingly color-coded in blue among all tasks within the cloud-shaped figure.  Memory controller (MC) nodes are not assigned any tasks as they contain MCs rather than PEs.


\subsection{Even Task Mapping}
\par DNN tiling strategies generally allocate an equal amount of work to each available resource, until the final mapping iteration for tail tasks, where the remaining tasks may be not enough for all PEs. Allocating tasks to the entire NoC at once constitutes one mapping iteration.  

\par
\begin{figure}[htb]
      \centering
    \includegraphics[trim=2cm 2.4cm 0cm 4.3cm, clip,width = 8 cm]{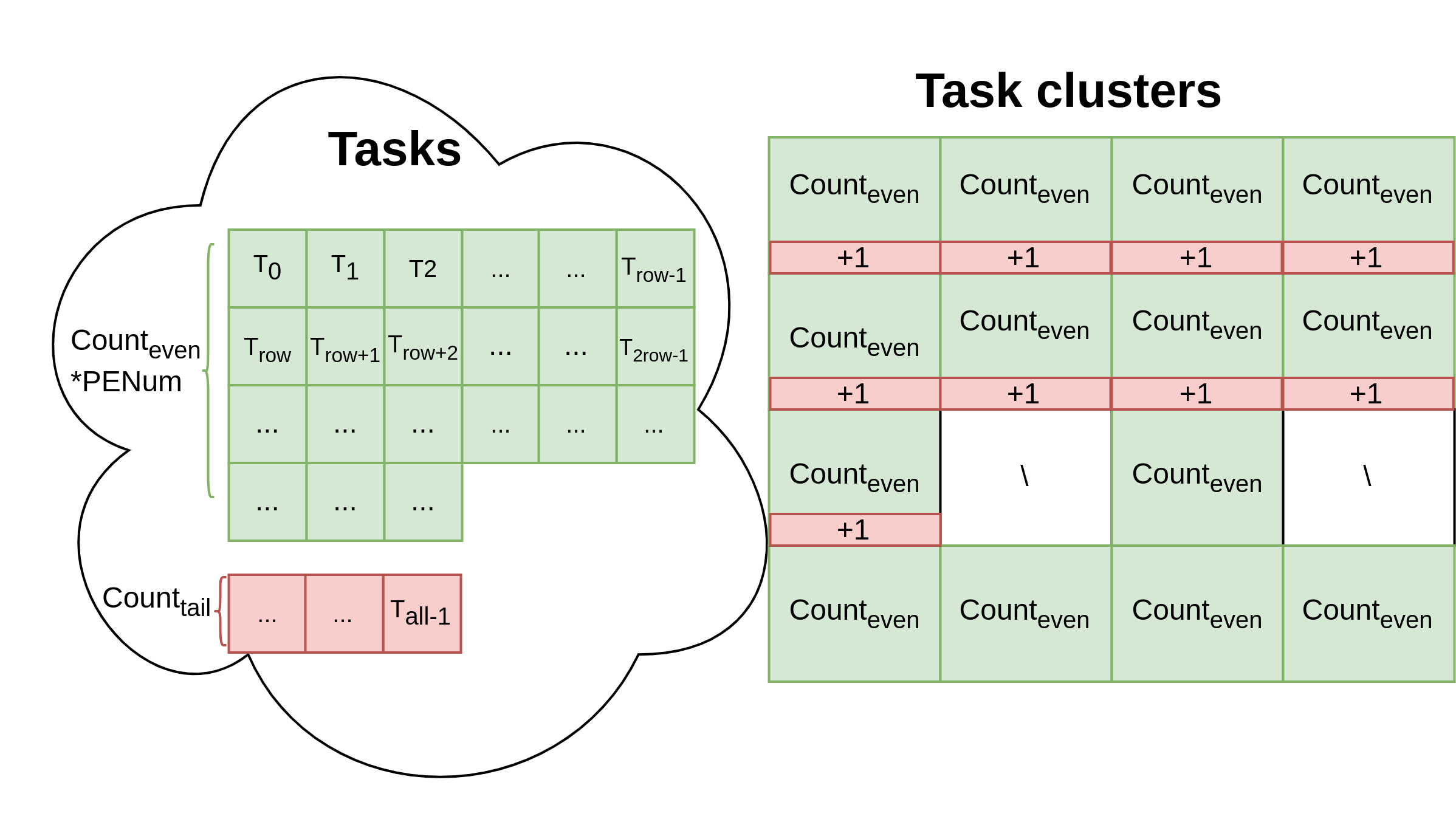}
      \caption{Even mapping}
      \label{fig:Even mapping}
  \end{figure}

  We present the row-major strategy in \fig{fig:Even mapping}.  The left side shows $Count_{\text{even}} \times \text{PENum}$ green tasks, which constitute the main portion, along with the tail tasks. These tasks are evenly mapped across the PEs following the row order.
\par However, even mapping lacks sensitivity to crucial information such as NoC architecture, and the real-time status of network congestion.

\subsection{Distance-Based Task Mapping}
\begin{figure}[htb]
      \centering
    \includegraphics[trim=0cm 0cm 3cm 2.3cm, clip,width = 11 cm]{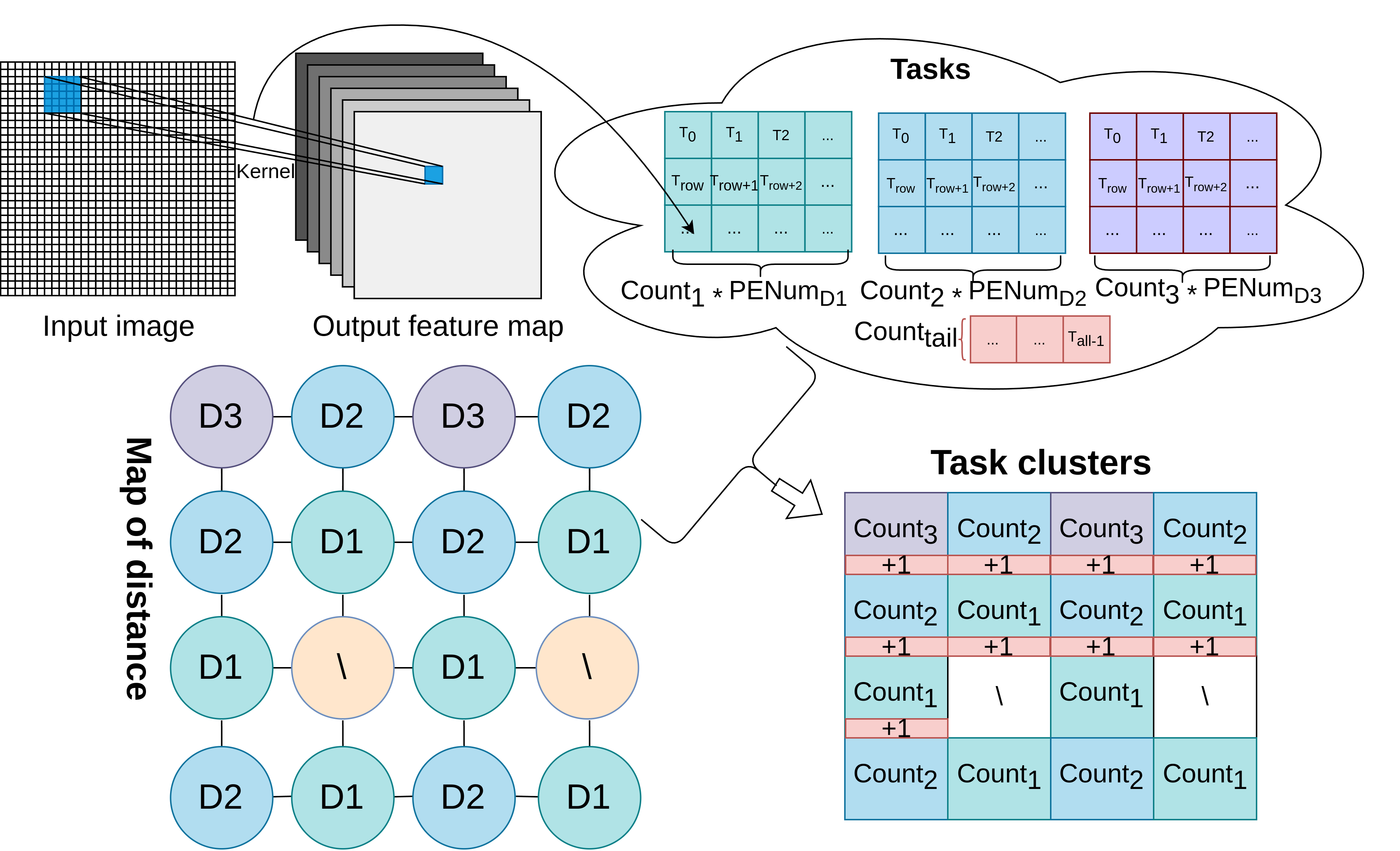}
      \caption{Distance-based mapping}
      \label{fig:distanceMapping}
  \end{figure}

To utilize the NoC architecture information, we explore the mapping according to different distances to memory nodes.  A node is categorized as "distance 1" if it directly surrounds an MC node, as shown in \fig{fig:distanceMapping}. Similarly, nodes further from the MC are classified as "distance 2" and "distance 3" and colored.

 Inspired by longer distances leading to longer time consumption, we distribute fewer tasks to far nodes as shown in \equ{distancetaskratio}. The total number of tasks is calculated by multiplying the number of nodes by the task count per node, as shown in \equ{distancetaskall}. Combining \equ{distancetaskratio} and \equ{distancetaskall}, we can compute the task number for each node and map these tasks accordingly.
\begin{equation}
Task_{count1}*Distance1 =Task_{count2}*Distance2= Task_{count3}*Distance3
\label{distancetaskratio}
\end{equation}
\begin{equation}
Task_{all} = Num_{D1}*Task_{count1} +Num_{D2}* Task_{count2} + Num_{D3}* Task_{count3}
\label{distancetaskall}
\end{equation}

\par  This approach relies on static distance information, leading to fixed ratios despite varying NN model configurations and run-time conditions.

\section{Travel time Based Mapping Approach}\label{Task mapping method}
\par We propose a travel time based mapping strategy for determining task allocation ratios in various scenarios. We explore the use of static latency, post-run recorded latency, and run-time travel time within a sampling window.
\subsection{Travel Time}
\par
\begin{figure}[htb]
      \centering
    \includegraphics[width = 6.5 cm]{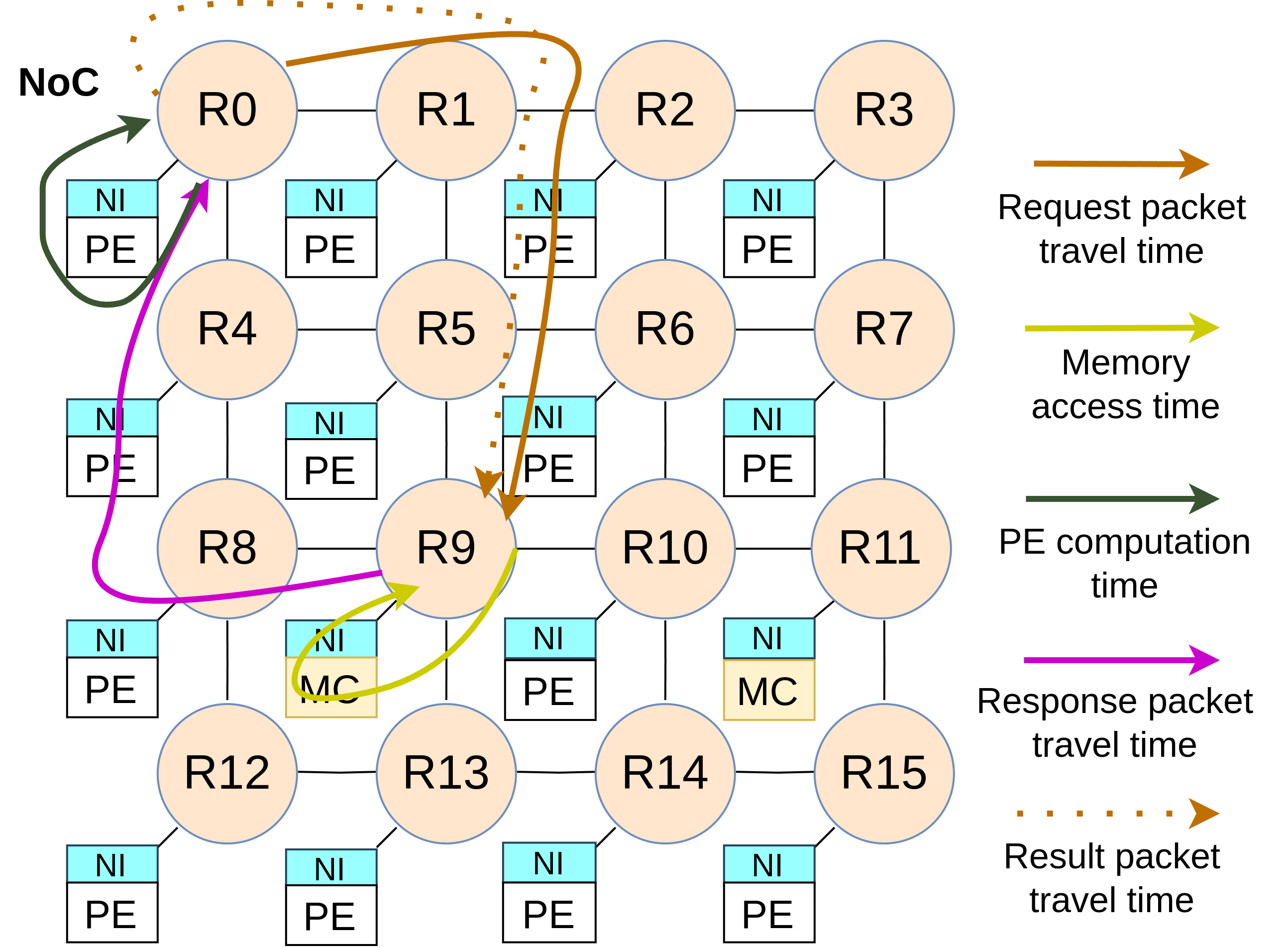}
      \caption{Travel time in NoC}
      \label{fig:travelTimeInNoC}
  \end{figure}

 As shown in \fig{fig:travelTimeInNoC}, travel time consists of different components.  The first step is to send a request packet from the PE node to the MC node to gather data as the brown path in \fig{fig:travelTimeInNoC}. This packet contains a compact payload such as the source ID and the index of the data needed, comprising only one single flit. 

\par There is a memory access delay for complex DNN models that have high data demands. We present the time as a light yellow path in \fig{fig:travelTimeInNoC}.
\par After picking up data, the MC node sends out a response packet, carrying the required inputs and weights, back to the PE node.  By dividing the packet size by the bit count of a single flit, larger kernels generate packets containing more flits, leading to a longer travel time in NoC. The trajectory is tracked from the moment the first flit leaves the MC node's NI until the last flit arrives at the requesting PE's router's VC buffer. This travel path is the purple path in \fig{fig:travelTimeInNoC}.

\par Computation time for one task is obviously important and it is determined by kernel sizes which specify the required operations and the available hardware resources.  Computation time, the dark green path in \fig{fig:travelTimeInNoC}, varies across different layers due to different kernel sizes but is constant in the same layer. 
\par The last step is to deliver the result packet to MC. PE will generate the next request packet while previous results are on the way. 
To highlight the feature of this overlap and avoid counting this overlapped travel time twice, we plot the result packet travel path by a dotted line.

We present the travel time used in our proposed mapping method in \equ{traveltimesum}. The NoC congestion status is not directly but implicitly included in  $T_{req}$ and $T_{resp}$.
\begin{equation}
T_{travel} = T_{req} +T_{memaccess}+ T_{resp} + T_{compu}
\label{traveltimesum}
\end{equation}

\subsection{Travel time Based Task Mapping}
\par
\begin{figure}[htb]
\centering
\includegraphics[width = 11 cm]{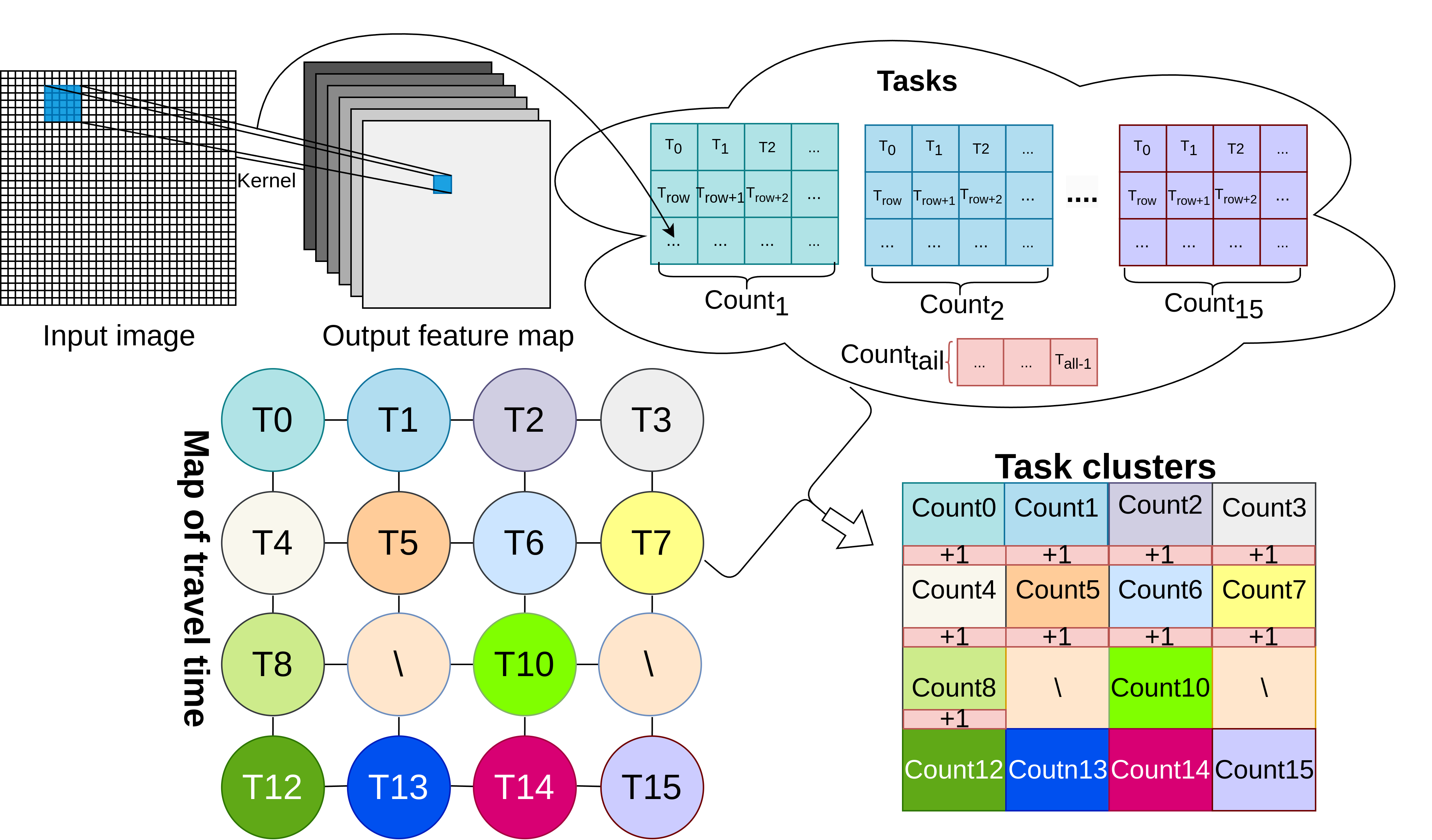}
\caption{Travel time based mapping}
\label{Travel time-based mapping}
\end{figure}

  Ideally, the overall latency is balanced by allocating different task counts in \equ{traveltimeratio}. The overall task count is shown in \equ{traveltaskall}.  By solving solve \equ{traveltaskall} and \equ{traveltimeratio}, we compute the task counts for the corresponding task cluster. Travel time Based mapping shown in \fig{Travel time-based mapping} is to allocate task clusters according to different PEs.  
\begin{equation}
\begin{split}
Task_{count1}*T_{travel1}=Task_{count2}*T_{travel2}=...= Task_{countN}*T_{travelN}
\end{split}
\label{traveltimeratio}
\end{equation}

\begin{equation}
Task_{all} =Task_{count1} + Task_{count2} +... +Task_{countN}
\label{traveltaskall}
\end{equation}

 \subsubsection{Static-Latency-Based Task Mapping}
\par  Without running on NoC, we can compute the static latency (SL) in \equ{equationstaticlatency} according to static and model information. The PE computation time is derived from the workload divided by the available MAC count.  The memory access time also depends on the workload and the bandwidth. By dividing the data size by bandwidth, we can compute the delay caused by memory access. Estimating time spent in NoC is challenging, but it will at least equal the product of distance and link delay. If multiple flits are involved, the latency between the head and tail flits must also be considered. Additionally, there are fixed overheads such as packetization latency.
\begin{equation}
T_{SL} = T_{compu}+T_{memaccess} +(D* T\_{link}+ (FlitNum-1)*T_{flit} )+ T_{fixed}  
\label{equationstaticlatency}
\end{equation}
\par


\subsubsection{Post-run Travel Time Based Mapping}
Travel time is precisely recorded during a complete run, and mapping is performed afterward. This method yields accurate travel time data for use in \equ{traveltimeratio}, but necessitates an additional run.
\subsubsection{Sampling Window in Travel Time Based Mapping}
\par
\begin{figure}[htb]
\centering
\includegraphics[width = 8 cm]{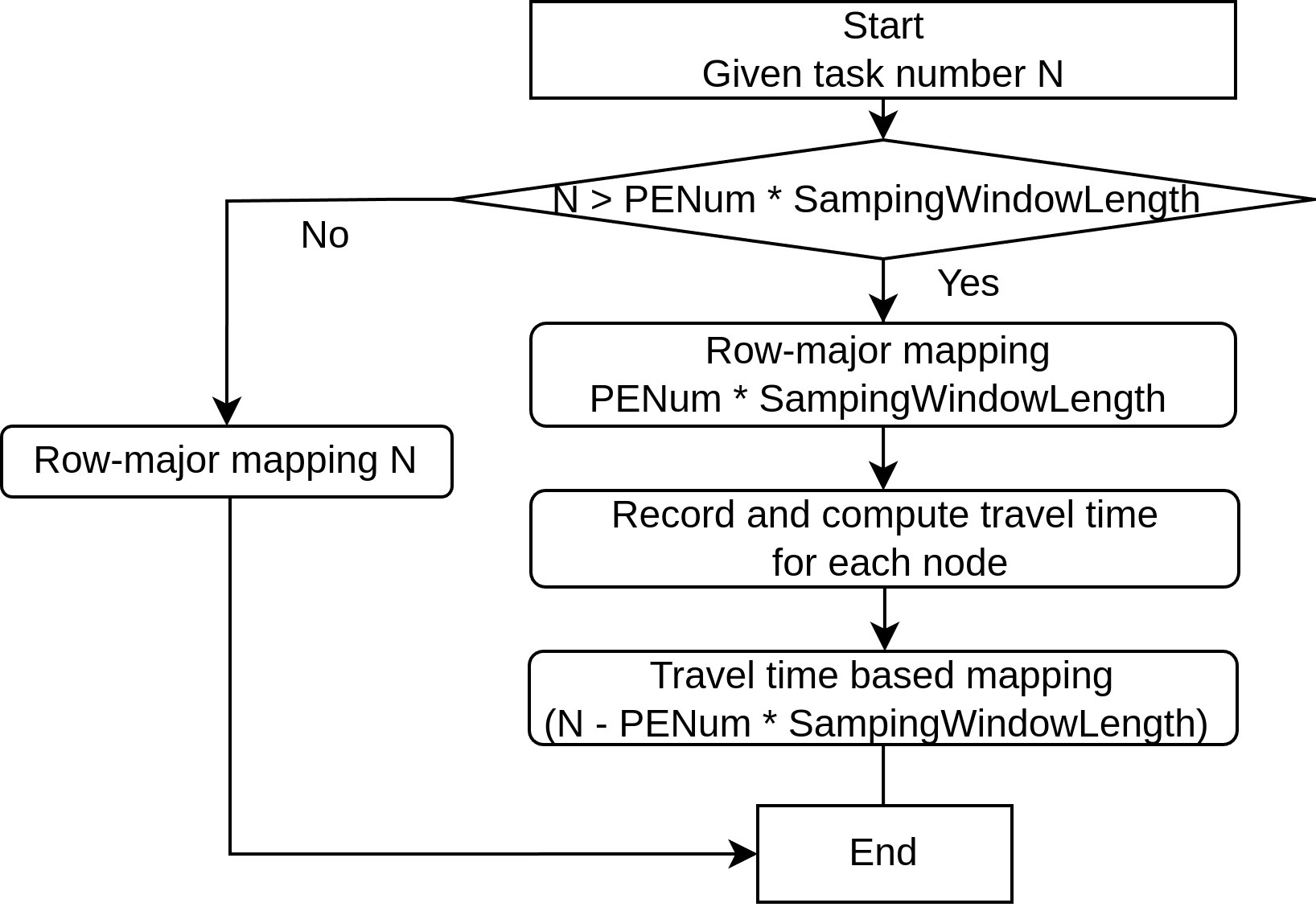}
\caption{Flowchart of travel time based task mapping with sampling window}
\label{fig:sampingWindowFlowChart.png2}
\end{figure}

  To do the mapping during runtime, we propose a travel time based mapping method utilizing a sampling window without an extra run. We divide the mapping process into two different routes in \fig{fig:sampingWindowFlowChart.png2}.  The left route is for a small layer without enough samples and a row-major mapping is directly used. For a layer with enough tasks, we employ a short sampling window to estimate travel times rather than recording accurate time after one complete run. The sampling window length, $SamplingWindowLength$, determines the sampled task number for one PE. After sampling, we recorded travel time and computed the task number to be allocated for each node.

\par This travel time obtained is not the precise travel time used in \equ{traveltimeratio}, but is the sampled time, $T_s$, in \equ{traveltimeratiosampled}. Additionally, the scope of task mapping shifts from encompassing all tasks to focusing on the residual tasks, as delineated in \equ{taskall-penum=taskformapping}. Combining  \equ{traveltimeratiosampled} and \equ{taskall-penum=taskformapping}, we can compute the task number allocated to each PE  after sampling.

\begin{equation}
Task_{count1}*T_{s1} = Task_{count2}*T_{s2}  = ...= Task_{countN}*T_{sN}
\label{traveltimeratiosampled}
\end{equation}
\begin{equation}
(Task_{all}- Task_{sampled}) =Task_{count1} + Task_{count2} +... +Task_{countN}
\label{taskall-penum=taskformapping}
\end{equation}

\section{Experiments}\label{sectexperiments}

\par  The performance difference in DNN accelerators could be modeled by three main factors: NN structure, communication protocol, and hardware platform. We abstract the difference between these three parts and configure three variables separately: mapping iteration, packet size, and NoC hardware architecture. 
\begin{itemize}
\item Mapping iterations/packet number: determined by different DNN input feature map sizes, DNN output channels, and DNN kernel sizes.
\item Packet size/flit number: determined by DNN kernel size and communication protocol which specifies the bits utilized by a single flit. 
\item NoC architecture: PE and MC nodes'  varying numbers and positions. 
\end{itemize}


\subsection{Experiment Setup}
\subsubsection{Simulation Platform}
First of all, we introduce the simulation platform, a cycle-accurate CNN-NoC accelerator simulation environment based on a behavior-level NoC simulator as outlined in \cite{wang2021flexible}. The NoC consists of a VC network derived from Gem5-Garnet \cite{agarwal2009garnet}, utilizing the widely used X-Y routing algorithm. Each physical connection between routers comprises four virtual channels, with each VC containing a four-flit buffer. The NoC's operating frequency is set at 2 GHz, following the specifications in \cite{wang2021flexible}, while the PEs are configured to function at 200 MHz, as detailed in \cite{hu2022high}. These configurations were selected due to their widespread acceptance in the field.

\par The nodes are labeled as PE nodes and MC (Memory controller) nodes, as shown in  \fig{fig:taskmapping_horizonal}. Each PE contains 64 Multiply-Accumulate (MAC) units, mirroring the 64 MACs per PE configuration in the Simba \cite{shao2019simba} architecture. The computation delay is calculated by dividing the required MAC operations by 64. For example, a convolution requiring 25 MAC operations consumes one PE cycle and 128 MAC operations consume 2 PE cycles. The two nodes connected to MC have a memory bandwidth of 64 GB/s, emulating a DDR5 device \cite{neda2023ciflow}. One data of weight or input is one 16-bit fixed point number which is two Bytes. Hence,  one data consumes $\frac{1}{32*10^9}s=0.0625$ router cycles, and the memory access delay is determined by the data number. 



 \par

\subsubsection{Configuration for Different Scenarios}
\par It is common that different models have varying input feature map sizes, kernel sizes, and output channels, affecting the number of tasks.  To investigate the impact of task count variations, we extend the task count with ratios from 0.5x  to 8x  by adjusting the output channel from 3 to 48, while the default configuration is 6 (1x). The first layer of LeNet contains an output of 6x28x28 corresponding to 4704 convolution tasks. This means 336 mapping iterations for even mapping due to 14 PE nodes. We explore a range of 2352 to 37632 tasks which means 168 to 2688 mapping iterations.

\par The convolution kernel size and the communication protocol's bit-per-flit configuration determine the flit number per task. As only the response packet contains data picked up from MC nodes, the request and result packets' sizes are not impacted.  We modify the kernel dimensions from 1x1 to 13x13 in \tab{Different kernel size and packet size}, including the original 5x5 kernel in the LeNet first layer. 

\begin{table}[]

\centering
\caption{Different kernel size and packet size}
\label{Different kernel size and packet size}
\begin{tabular}{|l|l|l|c|l|}
\hline
Input                  & Kernel size & Padding & \multicolumn{1}{l|}{Mapping iterations} & \begin{tabular}[c]{@{}l@{}}Packet size \\ in flits\end{tabular} \\ \hline
\multirow{6}{*}{28x28} & 1x1         & 0       & \multirow{6}{*}{336}                    & 1                                                               \\ \cline{2-3} \cline{5-5} 
                       & 3x3         & 1       &                                         & 2                                                               \\ \cline{2-3} \cline{5-5} 
                       & 5x5 (original)        & 2       &                                         & 4                                                               \\ \cline{2-3} \cline{5-5} 
                       & 7x7         & 3       &                                         & 7                                                               \\ \cline{2-3} \cline{5-5} 
                       & 9x9         & 4       &                                         & 11                                                              \\ \cline{2-3} \cline{5-5} 
                             & 11x11         & 5       &                                         & 16                                                            \\ \cline{2-3} \cline{5-5} 
                       & 13x13       & 6       &                                         & 22                                                              \\ \hline
\end{tabular}
\end{table}

Our investigation also extends to two different NoC architectures: one with two MCs and the other with four MCs.
\par We not only compare different methods but also evaluate the performance of our methods with different sampling windows. Unless otherwise specified, we use a sampling window of 10 in travel time-based DNN task mapping. 

\subsection{Experiment Results of Unevenness}
\subsubsection{Unevenness}
 We introduce the unevenness between task complementing time in \equ{eq1}. The unevenness $\rho$ is the difference between the maximum delay by the slowest node and the minimum latency by the fastest node divided by the maximum latency. Our analysis focuses on minimizing the maximum time consumption because it, rather than average time overhead, determines the final inference time for a layer. 

\begin{equation}\label{eq1}
    \rho = \frac{T_{max}-T_{min}}{T_{max}}
\end{equation}

\subsubsection{Unevenness under Row-major Mapping}
\par
\begin{figure}[htb]
 \begin{adjustwidth}{-3cm}{-3cm} %
\centering
\includegraphics[width=14cm]{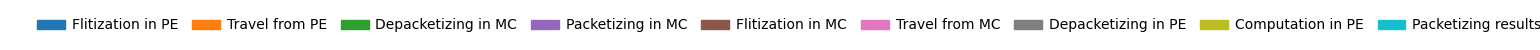}
\vspace{-2mm}
 
\centering
\begin{subfigure}{.31\textwidth}
  \centering
\includegraphics[width=3.8cm]{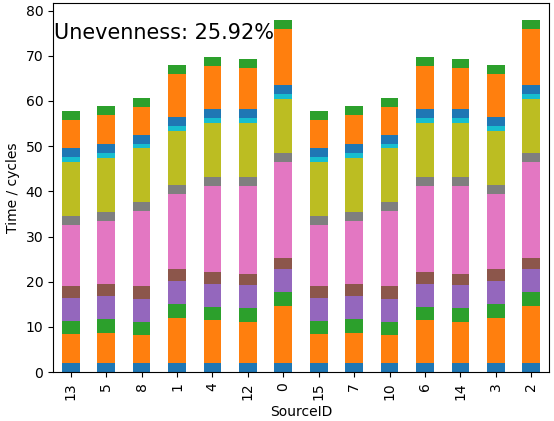}
\caption{}
\label{unevenxmappingavg.png}
\end{subfigure}%
\begin{subfigure}{.3\textwidth}
  \centering
\includegraphics[width=3.8cm]{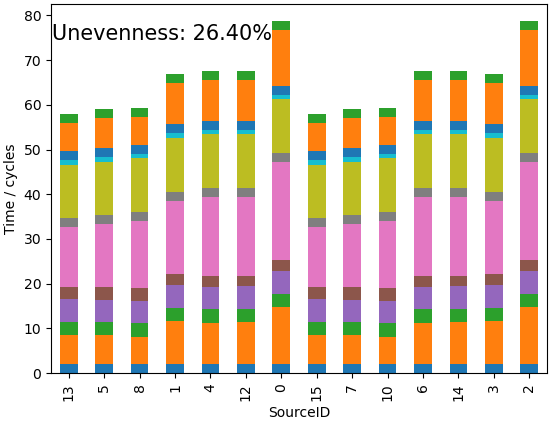}
\caption{}
\label{unevendistanceavg.png}
  \label{fig:sub2}
\end{subfigure}
\begin{subfigure}{.31\textwidth}
  \centering
\includegraphics[width=3.8cm]{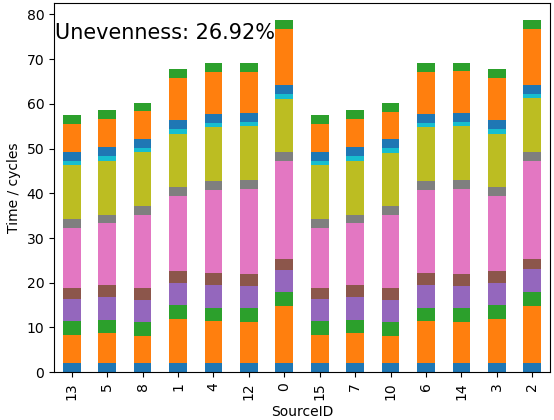}
\caption{}
\label{uneventravelavg.png}
\end{subfigure}
\begin{subfigure}{.3\textwidth}
  \centering
\includegraphics[width=3.8cm]{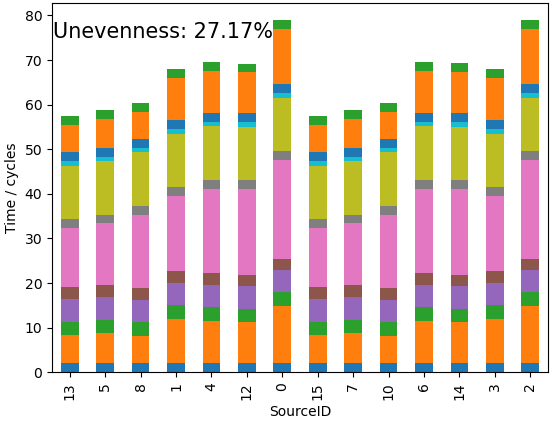}
\caption{}
\label{postuneventravelavg.png}
\end{subfigure}

\begin{subfigure}{.31\textwidth}
  \centering
\includegraphics[width=3.8cm]{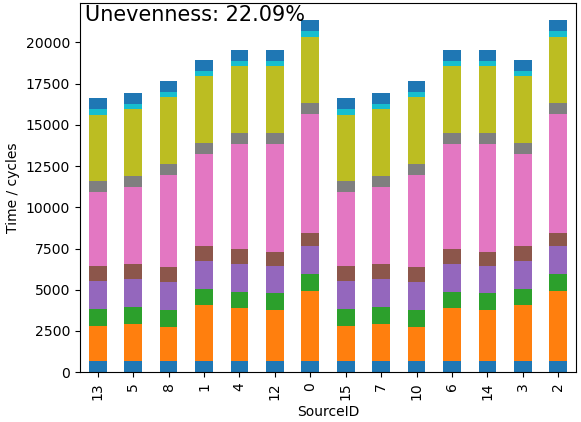}
\caption{}
\label{unevenxmappingsum.png}
\end{subfigure}
\begin{subfigure}{.3\textwidth}
  \centering
  \includegraphics[width=3.8cm]{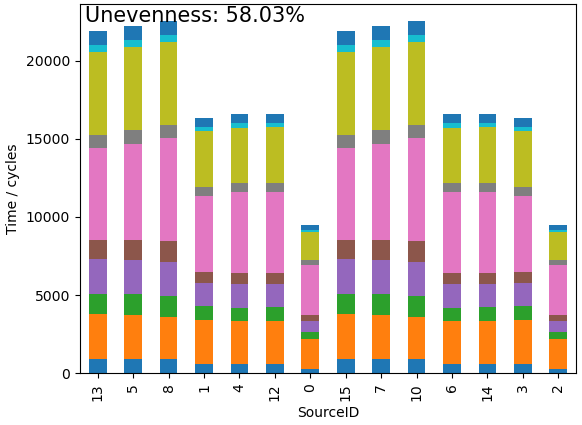}
\caption{}
\label{unevendistancesum.png}
\end{subfigure}
\begin{subfigure}{.3\textwidth}
  \centering
\includegraphics[width=3.8cm]{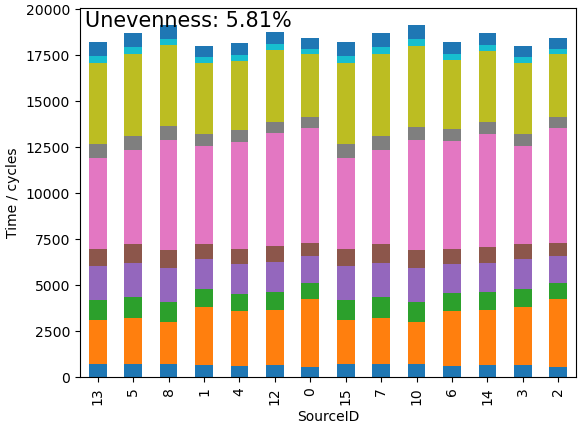}
\caption{}
\label{uneventravelsum.png}
\end{subfigure}
\begin{subfigure}{.3\textwidth}
  \centering
\includegraphics[width=3.8cm]{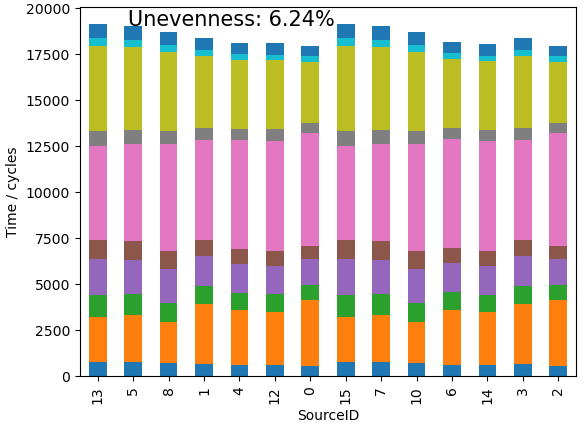}
\caption{}
\label{postuneventravelsum.png}
\end{subfigure}
\end{adjustwidth}
\caption{Results of unevenness (a)Average results for row-major mapping (b)Average results for distance-based mapping (c)Average results for travel time-based mapping (sampling window  10) (d)Average results for travel time-based mapping (post-run) (e)Accumulated results for row-major mapping (f)Accumulated results for distance-based mapping (g)Accumulated results for travel time-based mapping (sampling window  10) (h)Accumulated results for travel time-based mapping (post-run)}
\label{fig:test}
\end{figure}

 We present the unevenness in \fig{unevenxmappingavg.png}, showing the average time overhead for an end-to-end DNN task. To present the results more effectively, we arrange the results of the 14 PE nodes in an order of increasing distances. The end-to-end completion time for one DNN task ranges from 57.69 to 77.88 cycles, showing an unevenness of 25.92\%. 

\par Due to the overlap, we present partially accumulated time consumption in \fig{unevenxmappingsum.png}, where the travel times of result packets and subsequent delays are excluded.  Within the same PE, a response packet is always generated after the request packet. Although packets from different PEs may interleave within the network, they are processed sequentially within their respective PEs. Consequently, we depict them stacked in Figure \ref{unevenxmappingsum.png}.
The unevenness in the partially accumulated time consumption is 22.09\%, showing that the fastest PE—represented by the lowest bar in \fig{unevenxmappingsum.png}—is idle for 22.09\% time in this layer while the slowest PE is continuing its operation. This scenario highlights the inefficiency in resource usage and underscores the necessity for uneven mapping strategies.

\subsubsection{Unevenness in Distance-Based Mappings}
\par From \fig{unevendistanceavg.png}, we observe that travel time is significantly influenced by the NoC architecture, more exactly, the distances. Nodes 13, 5, and 8 are the fastest as they have only one distance, and nodes 1, 4, and 12 consume moderate time as their distances are two. Node 0 has the longest distance, three, and requires the most time.  However, simply allocating the task following the ratio of distance, leads to the unevenness of 58.03\% in \fig{unevendistancesum.png}. This reveals that we need a better mapping method to set the ratio rather than simply adopting the distances as ratios.    


\subsubsection{Unevenness in Travel Time-Based Mapping}
\par The outcomes of travel time-based mapping with a sampling window of 10 tasks in \fig{uneventravelavg.png} and post-run travel time-based mapping in \fig{postuneventravelavg.png}, also identify three stages of travel times across different PEs, which are associated with distance groups. By allocating a variable number of tasks to each PE, we nearly equalize the total time consumption for all packets per PE in \fig{uneventravelsum.png} and \fig{postuneventravelsum.png}.  We effectively balance overall time consumption on each PE and achieve an unevenness of 5.81\% and 6.24\%.

\subsection{Results for Different Mapping Iterations}
\par
\begin{figure}[htb]
      \centering
    \includegraphics[trim=3cm 0cm 4.5cm 2.5cm, clip,width = 12 cm]{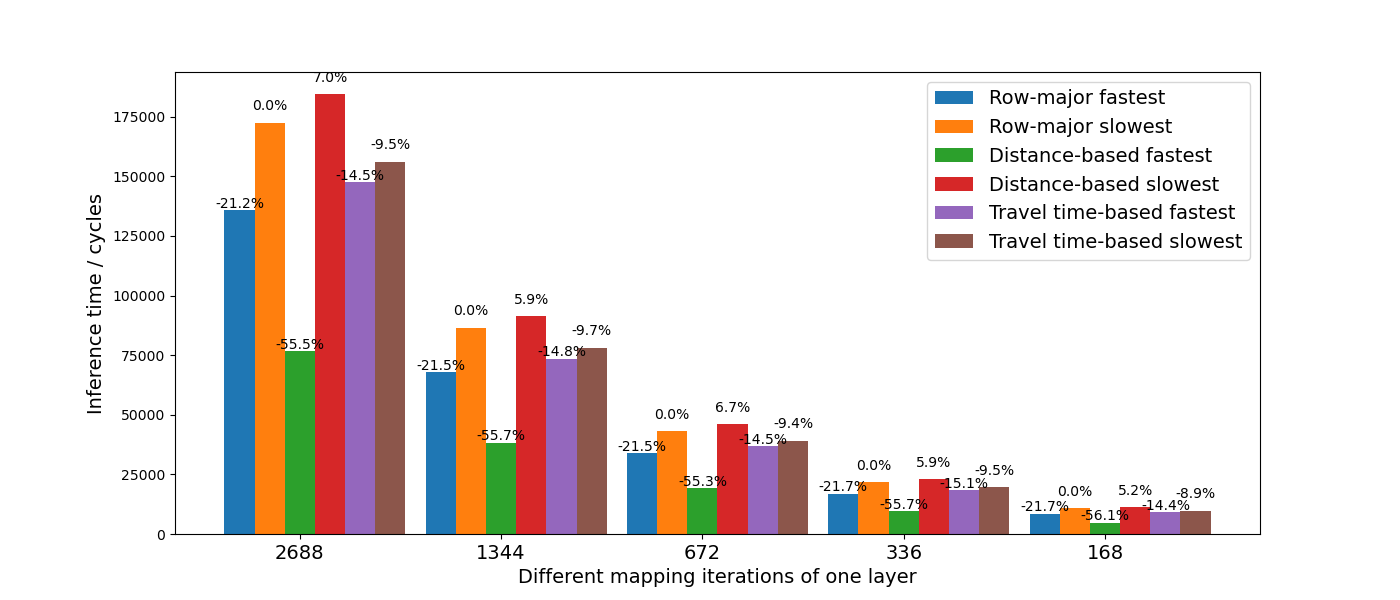}
 \caption{Different mapping iterations}
  \label{mappingiterations}
  \end{figure}

 We present the inference time of different mapping iterations in \fig{mappingiterations}. To better compare the time consumption across different mapping methods, we show the percentage difference of each bar to the orange bar, which is the slowest PE in row-major mapping.  For row-major mapping, under all mapping iterations, there is a gap of around 21\% and this shows potential to improve as one PE is idle while the other PE is working. In distance-based mapping, the fastest PE operates for a shorter duration compared to its counterpart in row-major mapping, while the slowest PE works for a longer period. This outcome illustrates the challenge and complexity of achieving effective uneven mapping.

\par Our proposed travel time mapping significantly narrows the gap between the low bar and high bar in the same mapping method from 21.2\% to around 5\%. Compared with row-major mapping, the total time consumed for processing a single layer is dominated by the slowest PE, and our method achieves notable improvement around 9.7\%.

\subsection{Results for Different Packet Sizes}

 \par
\begin{figure}[htb]
\centering
\includegraphics[trim=2cm 0pt 2cm 2cm, clip, width = 12 cm]{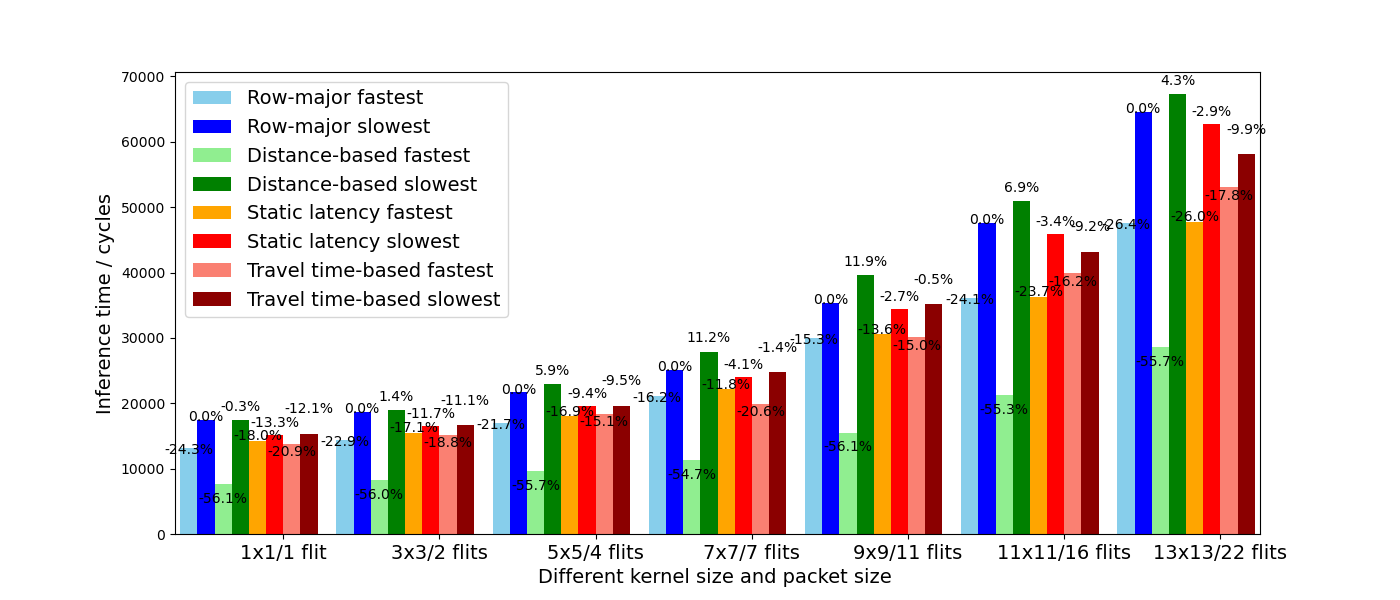} 
\caption{Inference time for one layer with varying kernel size and packet size}
\label{fig:flitnumberresults}
\end{figure}

 \fig{fig:flitnumberresults} reveals that unevenness exists across different packet sizes. All distance-based mapping worsens the situation by increasing the unevenness together with the final latency dominated by the slowest PE. Static-latency-based mapping achieves high performance while the flit number is small. However, as the number of flits increases, more flits are injected into the NoC, and the exclusion of congestion and queuing delays in the static latency calculations leads to decreased performance. Our proposed method enhances the latency for one single layer, reducing the time consumption of one layer up to 12.1\%.   This demonstrates the complexity of setting the ratio for uneven mapping and validates the effectiveness of our proposed method. 


\subsection{NoC Architecture}
\label{sect:NoC Architecture}
\par
\begin{figure}[htb]
 \begin{adjustwidth}{-1cm}{-1cm} 
\centering
\begin{subfigure}{.2\textwidth}
  \centering
\includegraphics[width=2.5cm]{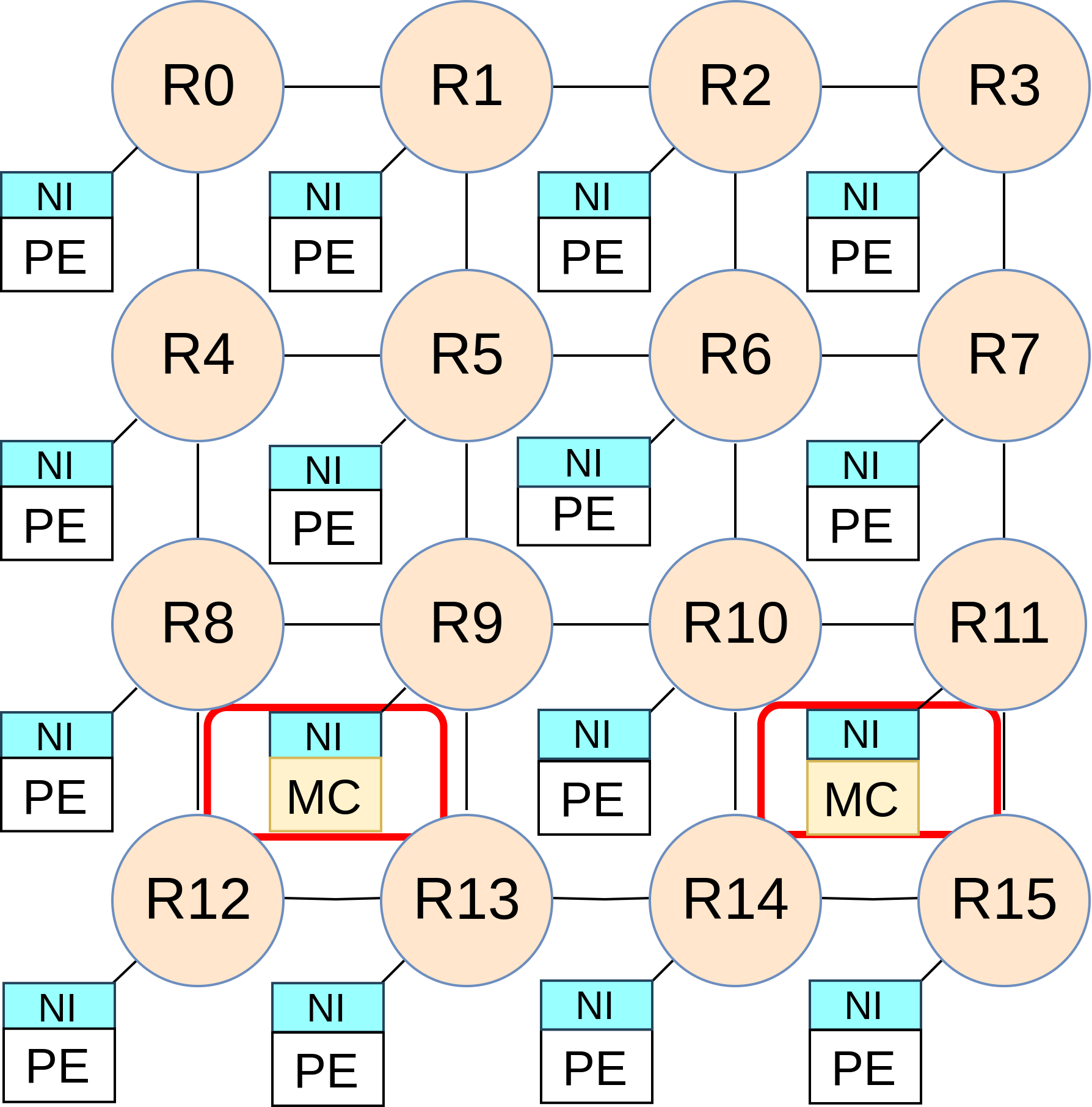}
\caption{4x4 NoC with two MC nodes}
\label{(a)4x4 NoC with two MC nodes}
\end{subfigure}%
\begin{subfigure}{.2\textwidth}
  \centering
\includegraphics[width=2.5cm]{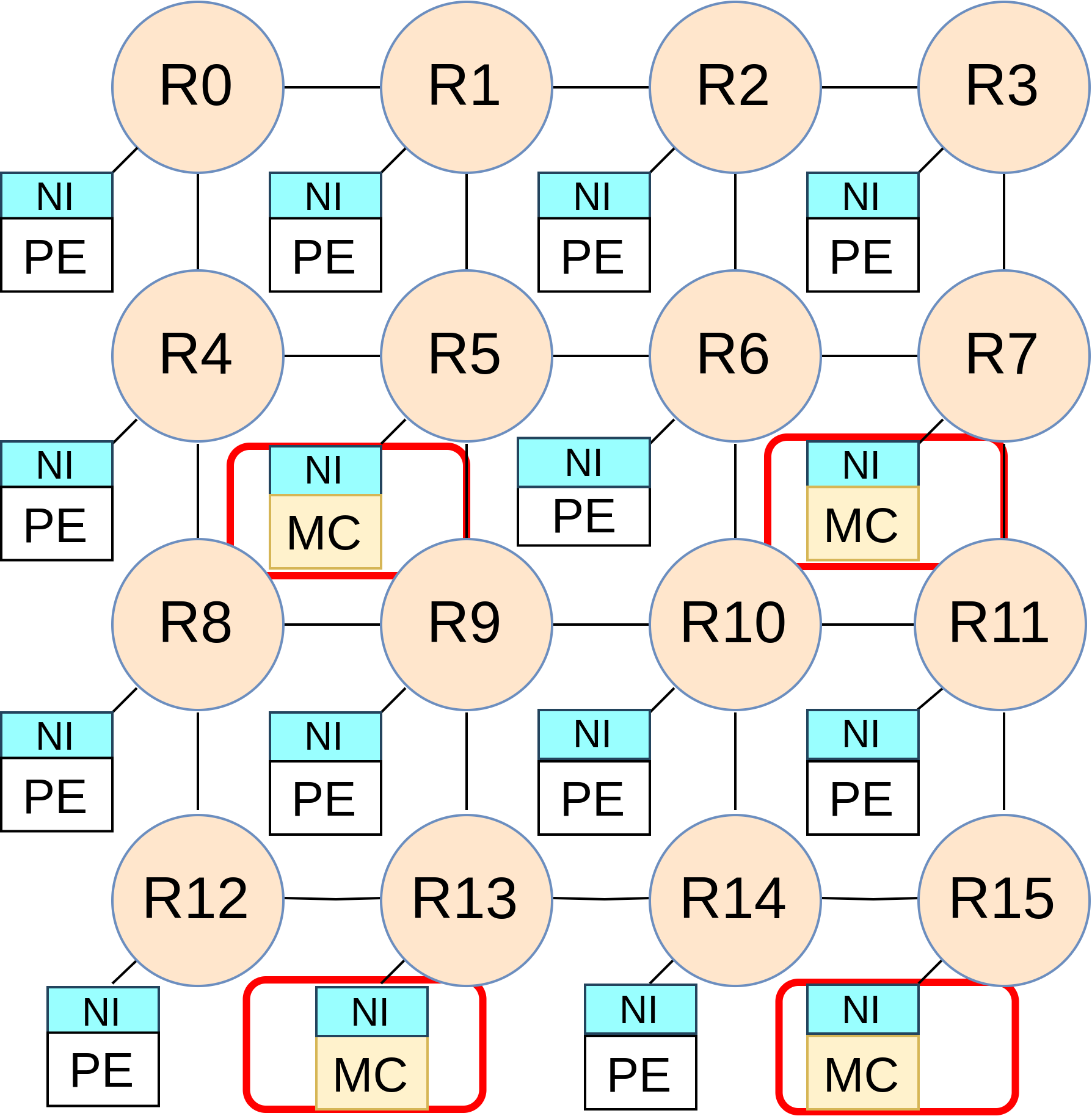}
\caption{4x4 NoC with four MC nodes}
\label{(b)4x4 NoC with four MC nodes}
\end{subfigure}
\begin{subfigure}{.6\textwidth}
\centering
\includegraphics[trim= 3cm 0cm 3cm 2.5cm, clip,width=8.2cm]{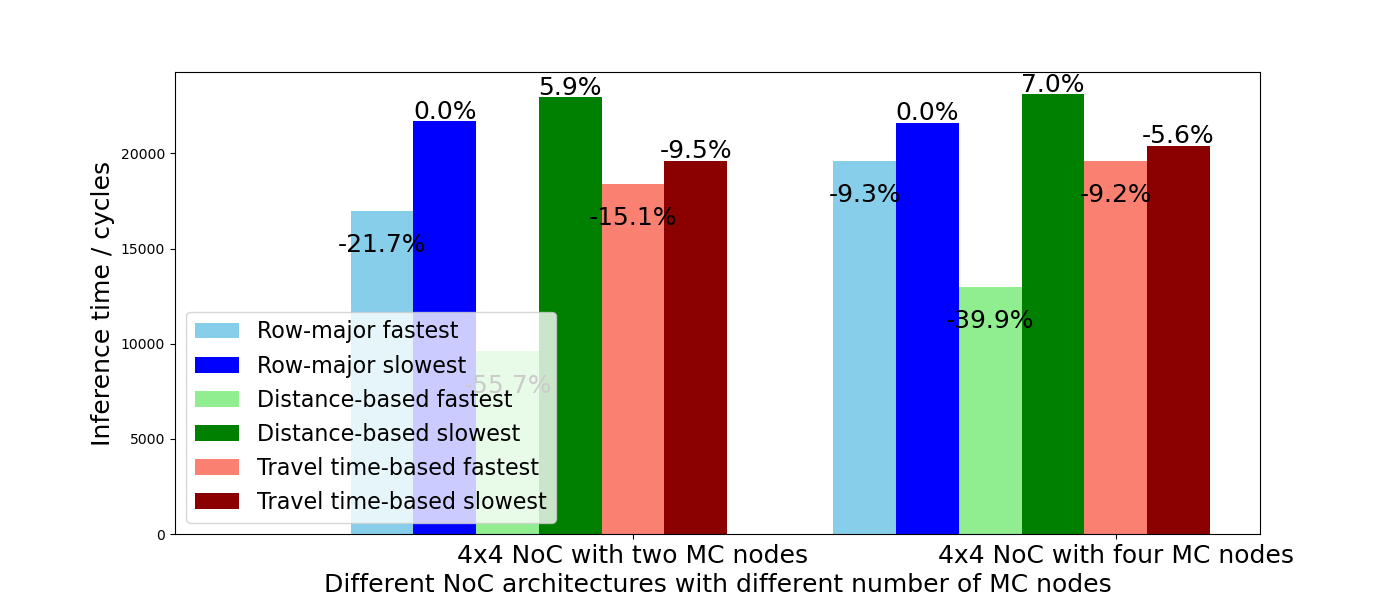} 
\caption{Inference time under two NoC architecture}
\label{(c) Inference time of one layer under two NoC architecture}
\end{subfigure}
\caption{Results of different NoC architectures}
\label{fig:different-Noc-architecture}
\end{adjustwidth}
\end{figure}

 We present different NoC architectures in   \fig{(a)4x4 NoC with two MC nodes} and \fig{(b)4x4 NoC with four MC nodes} while one of them contains two MC nodes and one of them contains four MC nodes. Compared to row-major mapping, the enhancement provided by our method in configurations with four MC nodes is 5.6\%, which is less than the 9.5\% improvement observed in the default settings. This reduced enhancement is caused by the minimal variance in distances among nodes, which lowers the gap between the fastest PE and the slowest PE in row-major mapping from 21.7\% to 9.3\%, resulting in a narrower space for optimization.

\vspace{-2mm}
\subsection{Experiment Results of Mapping Whole NN Model with Different Sampling Windows}
\label{Experiment Results of Mapping NN Model with Different Sampling Windows}
\par
\begin{figure}[htb]
 \begin{adjustwidth}{-1cm}{-1cm} 
\centering
\includegraphics[width = 12 cm]{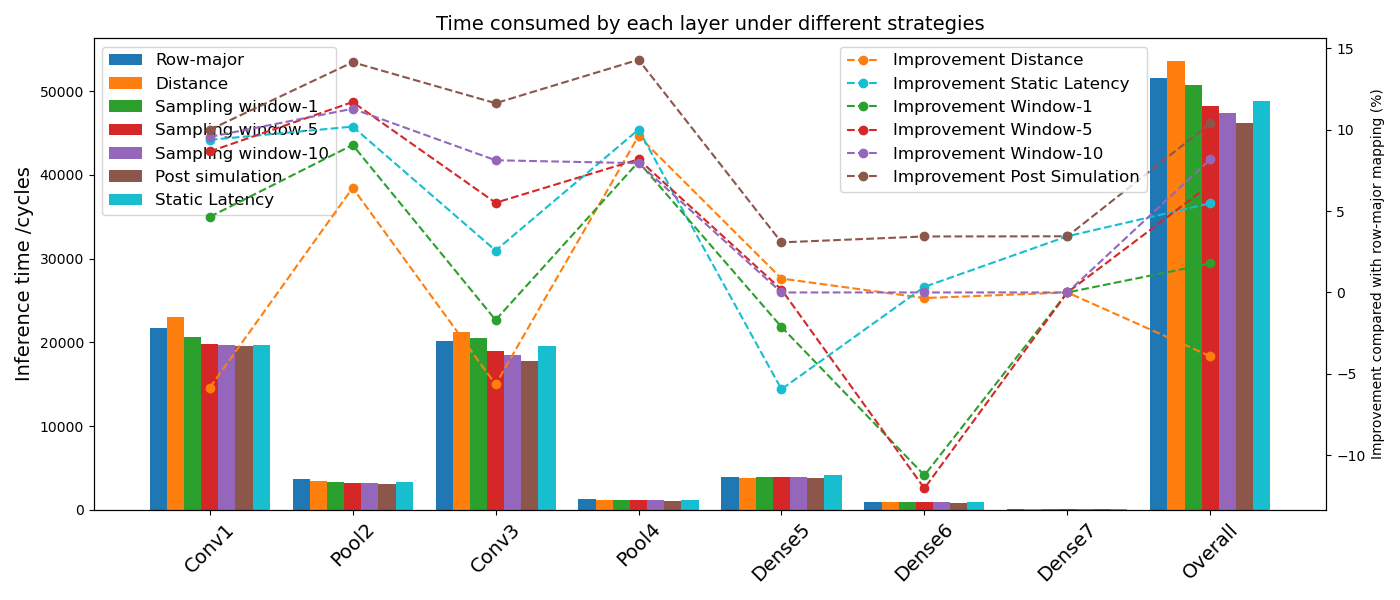}
\caption{Inference time for LeNet}    
\label{fig:Inference time for LeNet}
\end{adjustwidth}
\end{figure}

   We present the inference times across eight clusters, each representing one of the 7 individual layers and an aggregate of the overall results in \fig{fig:Inference time for LeNet}. Within each cluster, six distinct inference times are documented: those resulting from row-major mapping, distance-based mapping, and travel time based mapping with sampling windows of 1, 5, and 10, alongside the post-run travel time based mapping. The percentage improvements over row-major mapping, for each cluster, are also presented in polylines, offering a comprehensive comparison of each mapping strategy under varying conditions.

\par When comparing different mapping methods, with row-major mapping serving as the baseline, distance-based mapping sometimes outperforms and at other times gets worse results than row-major mapping across different layers. Static latency based mapping achieves a performance level that is intermediate between travel time based mapping of long and short sampling windows. For a sampling window of 1, performance drops are observed in layers 3, 5, and 6 due to the extremely few sampling packets. With a sampling window of 5, the performance is improved and only layer 6 shows a decrease compared to row-major mapping. It may be caused by the small packet count of 84 in layer 6; the minor loss is just 105 cycles. With a longer sampling window of 10, the performance no longer worsens compared to row-major mapping in any layer.

All travel time based mappings show improvement over row-major mapping and over distance-based mapping for the whole time consumption. With an increased sampling window length from 1 to 10, the overall improvement increases from 1.78\% to 8.17\%, approaching the ideal post-run travel time based mapping of 10.37\%.


\subsection{Summary and Discussion}

\par We summarize the experiment results as itemized below:  
\begin{itemize}
 
\item Unevenness exists in even mapping, across various configurations including mapping iterations, packet size, and NoC architecture. Our proposed mapping consistently outperforms both row-major and distance-based mapping by dynamically determining the task ratio according to the travel time. 
   \item A post-run travel time based mapping invariably offers the best performance in all layers and the whole model, compared with row-major mapping, distance-based mapping, and travel time based mapping with a sampling window. However, it needs one additional run to collect accurate travel time, resulting in extra time and energy consumption.
   \item Utilizing a sampling window during operation time provides sampled rather than precise travel times for mapping, eliminating the need for an extra run. Larger windows tend to offer better approximations as more samples are recorded. A sampling window of 10 offers an effective trade-off between performance and sample number, achieving an 8.17\% improvement for a whole NN model, closely approaching the ideal post-run gain of 10.37\%.
\end{itemize}

\par The superior results of post-run travel time based mapping are derived from three main factors: 1) Enhanced accuracy as it tracks all packets rather than several samples; 2) Ability to work in smaller layers lacking sufficient tasks for sampling; 3) Unlike adjusting ratio after sampling, which benefits only leftover tasks, post-run mapping benefits all tasks.
\par The choice of sampling window length is crucial.      For a window length of 1, the travel time has a bias compared to the accurate travel time in post-run, leading to varying final delay improvement for the whole LeNet compared with row-major mapping, from 1.78\% for window 1 and 10.37\% for post-run. This indicates that a too-small sample size is ineffective. Compared to the 10.37\% improvement achieved by the ideal post-run outcomes, our implementation of a 10-task sampling window achieves an 8.17\% improvement over row-major mapping. This high effectiveness is due to the repetitive nature of NN tasks, which makes network activities similarly predictable and minimizes the need for extensive sampling.

\section{Conclusion}
\label{sectconclusion}
\par  We have proposed a dynamic, travel time based mapping strategy to mitigate the issue of uneven time consumption across PEs due to imbalanced task mapping, which leads to inferior performance. We compare our approach with traditional even mapping and distance-based mapping strategies. The results indicate that relying solely on the distance information is inadequate, thereby validating the significance of our method. Additionally, we examine the effects of different mapping iterations, packet sizes, and NoC architectures—factors corresponding to varying neural network models, communication protocols, and hardware platforms. Beyond collecting information from extra running, we proposed a technique for sampling travel time information during runtime and evaluated the impact of different window lengths.

\par The findings reveal a 22.09\% unevenness in row-major mapping, which our method reduces to 5.81\%. Consequently, we achieved a maximum 9.7\% improvement in latency for varying mapping iterations, 12.1\% for different packet sizes, and 9.7\% for diverse NoC architectures.  With post-run travel time based mapping, our method obtains an enhancement of 10.37\% for the whole LeNet. By implementing dynamic window lengths, we were able to estimate improvements without additional simulations, achieving 1.78\%, 6.62\%, and 8.17\% improvements for sampling window lengths of 1, 5, and 10, respectively. This reveals that using more samples achieves a performance more similar to the ideal result with post-run.

\par In our future work, our proposed approach should be compared with other adaptive NoC mapping approaches to assess the power and area overheads. In addition, since wiring and thermal management will directly affect the NoC performance, their impact when applying our solution is worth investigating.

\bibliographystyle{splncs04}
\bibliography{mybibliography}

\end{document}